\documentclass{PoS}
\usepackage{epsfig}
\usepackage{cite}
\newcommand{\comment}[1]{}
\usepackage{lipsum}
\usepackage{enumitem}

\usepackage{graphicx}
\usepackage{epsfig}
\usepackage{bm}
\usepackage{latexsym,amssymb,amsmath,amsfonts,amssymb,txfonts,pxfonts,wasysym,float}
\usepackage{color}

\DeclareMathOperator*{\argmax}{arg\,max}

\DeclareBoldMathCommand\blpar{\left(} 
\DeclareBoldMathCommand\brpar{\right)}

\def\bpar#1{%
\boldsymbol{\left(\vphantom{#1}\right.\mspace{-2mu}}%
#1%
\boldsymbol{\mspace{-2mu}\left.\vphantom{#1}\right)}}

\newcommand{\beq}[1]{\begin{equation}\label{#1}}
\newcommand{\eeq}{\end{equation}}
\newcommand{\bea}[1]{\begin{eqnarray} \label{#1}}
\newcommand{\eea}{\end{eqnarray}}
\newcommand{\ba}{\begin{array}}
\newcommand{\ea}{\end{array}}

\def\be{\begin{equation}}
\def\ee{\end{equation}}
\def\gs{\mathrel{
   \rlap{\raise 0.511ex \hbox{$>$}}{\lower 0.511ex \hbox{$\sim$}}}}
\def\ls{\mathrel{
   \rlap{\raise 0.511ex \hbox{$<$}}{\lower 0.511ex \hbox{$\sim$}}}}

\def\ie{i.e.\ }

\newcommand{\postscript}[2]{\setlength{\epsfxsize}{#2\hsize}
   \centerline{\epsfbox{#1}}}

\usepackage[usenames,dvipsnames]{xcolor}
\definecolor{orange}{cmyk}{0,0.5,1,0}
\definecolor{rossoCP3}{cmyk}{0,.88,.77,.40}
\definecolor{graa}{rgb}{0.8,0.8,0.8}
\definecolor{blaa}{rgb}{0.2,0.2,0.6}

\title{Evidence for UHECR origin in starburst
  galaxies}

\ShortTitle{Evidence for UHECR origin in starburst  galaxies}

\author{Luis A. Anchordoqui\\
Department of Physics \& Astronomy,  Lehman College, CUNY, NY 10468, USA\\
Department of Physics,
 Graduate Center, City University
  of New York,  NY 10016, USA\\
Department of Astrophysics,
 American Museum of Natural History, NY
 10024, USA\\
        E-mail: \email{luis.anchordoqui@gmail.com}}

\author{\speaker{Jorge F. Soriano} \\
Department of Physics \& Astronomy,  Lehman College, CUNY, NY 10468, USA\\
Department of Physics,
 Graduate Center, City University
  of New York,  NY 10016, USA\\
E-mail: \email{jfdezsoriano@gmail.com}}

      \abstract{The quest for the origin(s) of ultra-high-energy 
  cosmic rays (UHECRs) continues to be a far-reaching pillar of high
  energy astrophysics. The source scrutiny is mostly based
  on three observables: the energy spectrum, the nuclear composition,
  and the distribution of arrival directions. We
  show that each of these three observables can be well reproduced with
  UHECRs originating in starburst galaxies. }

\FullConference{36th International Cosmic Ray Conference - ICRC2019 -\\
		July 24 - August 1, 2019\\
	 Madison, Wisconsin, USA}

\begin{document}

Starburst galaxies are observed to be forming stars at an unusually
fast rate (about $10^3$ times greater than in a normal galaxy). The
areas of high activity can be spread throughout the galaxy, but most
star forming regions are observed in a small sector around the
nucleus. The starburst activity usually drives galactic-scale outflows
or ``superwinds'' that may be responsible for removing metals from the
galactic disk and polluting the intergalactic medium with
ultra-high-energy ($E \gtrsim 10^{9}~{\rm GeV}$) cosmic ray (UHECR)
nuclei~\cite{Anchordoqui:1999cu,Anchordoqui:2002dj,Anchordoqui:2018vji,Anchordoqui:2018qom}. Starburst
superwinds are powered by massive star winds and by core collapse
supernovae which collectively create hot ($T \lesssim 10^8~{\rm K}$)
bubbles of metal-enriched plasma within the star forming regions~\cite{Heckman}. The
over-pressured bubbles expand, sweep up cooler ambient gas, and
eventually blow out of the disk into the halo, providing a profitable
arena for the formation of collisionless plasma shock waves, in which
UHECRs can be accelerated by bouncing back and forth across the
shock. Herein we present additional support for this idea by
confronting the predictions of the model with experimental data.

Specific assumptions are made, in that we consider diffusive shock
acceleration on a distribution of particles at multiple parallel
shocks (in which both the magnetic field and the upstream and
downstream plasma flows are always perpendicular to the plane of the
shock front).\footnote{It has long been known that stellar winds
  contain a network of embedded shocks~\cite{White}. This provides
  some support for our conjecture.} Note since the magnetic field has
components only along the direction in which the shock propagates the
Rankine-Hugoniot jump conditions are hydrodynamic in character (see
Appendix). At each shock a new distribution of particles is injected
and accelerated, and the particles injected at earlier shocks are
re-accelerated further. Adiabatic decompression occurs after each
shock. We show that these considerations reduce the time constraint on
the acceleration region, while addressing the criticism on the model
raised in~\cite{Matthews:2018laz}.  Moreover, the presence in the wind
of many shocks changes the particle spectrum from that produced by a
single shock~\cite{Schneider}. Summing over an infinite number of
identical shocks, with fresh injection at each shock and decompression
between the shocks, does produce a power-law momentum distribution
$f_\infty (p) \propto p^{-3}$~\cite{Pope}, which is flatter than that
produce by a single shock $f(p) \propto p^{-4}$, and better reproduce
observations.

The UHECR spectrum can be roughly described by a twice-broken power
law~\cite{Abbasi:2007sv,Abraham:2008ru,Abraham:2010mj,AbuZayyad:2012ru}. The
first break is a hardening of the spectrum, known as ``the ankle.''
The second is an abrupt softening of the spectrum, which {\it (i)}~may
be interpreted as the long-sought GZK
cutoff~\cite{Greisen:1966jv,Zatsepin:1966jv}, or {\it (ii)}~may
correspond to the ``end-of-steam'' for cosmic
accelerators~\cite{Allard:2008gj,Aloisio:2009sj}. Herein we introduce
a complementary explanation {\it (iii)}~in which GZK interactions at
the source constrain the maximum energy of the nuclei. Note that {\it
  (iii)} is markedly different from {\it (ii)} because for a nucleus
of charge $Ze$ and baryon number $A$, the maximum energy of
acceleration capability of the sources grows linearly in $Z$, while
the energy loss per distance traveled decreases with increasing
$A$. The ankle energy and the corresponding change in the power-law
spectral index are measured with high precision. The existence of the
flux suppression is also firmly established. The differential energy
spectra measured by the Telescope Array (TA)
experiment~\cite{Abbasi:2007sv,AbuZayyad:2012ru} and the Pierre Auger
Observatory (Auger)~\cite{Abraham:2008ru,Abraham:2010mj} agree within
systematic errors below $E \sim 10^{10}~{\rm GeV}$; at higher
energies, TA observes more cosmic rays than would be expected if the spectral shape were the same as that seen by Auger. The flux
suppression observed in Auger data is at $E \sim 10^{10.6}~{\rm GeV}$,
whereas the one observed in TA data is at
$E \sim 10^{10.73}~{\rm GeV}$. 

The TA Collaboration has interpreted their data as implying a
  light primary composition (mainly $p$ and He) from $10^{9.1}$ to
  $10^{10.6}~{\rm GeV}$~\cite{Abbasi:2014sfa,Abbasi:2018nun}. The
  Auger Collaboration, using post-LHC hadronic interaction
  models, reports a composition becoming light up to $10^{9.3}~{\rm
    GeV}$ but then becoming heavier above that energy, with the mean
  mass intermediate between protons and iron at $10^{10.5}~{\rm
    GeV}$~\cite{Abraham:2010yv,Aab:2014kda,Aab:2014aea,Aab:2017cgk,Aab:2016htd}. Auger
  and TA have also conducted a thorough joint analysis and state that,
  at the current level of statistics and understanding of systematics,
  both data sets are compatible with being drawn from the same parent
  distribution, and that the TA data is compatible both with a
  protonic composition below $10^{10}~{\rm GeV}$ and with the mixed
  composition above $10^{10}~{\rm GeV}$ as reported by
  Auger~\cite{Abbasi:2015xga,Hanlon:2018dhz}. However, Auger data are
  more constraining and not compatible with the pure protonic option
  available with TA alone. Moreover, a recent re-analysis of TA
  data seems to indicate that a pure proton composition above the ankle
  is disfavored~\cite{Watson:2019clu}. 

  The high frequency spectral fall-off and the shape of the spectrum
  at and below the corner frequency are  critical to assess the
  characteristics of the source spectra. In particular, a
  simultaneous fit to the spectrum and the elongation rate requires
  hard source spectra $\propto E^{-\gamma}$, with
  $1.0 \lesssim \gamma \lesssim
  1.5$~\cite{Aab:2016zth,Unger:2015laa}. The differential energy
  spectrum $dN/dE \propto E^{-\gamma}$ is related to the phase space
  distribution in momentum space by $dN = 4 \pi p^2 f_\infty(p) \, dp$,
  yielding good agreement with Auger data. The constraint on the source
  spectral index would be relaxed if the number of UHECR sources
  increases at low redshifts (for such an unusual redshift evolution
  softer source spectra with $dN = 4 \pi p^2 f(p) \, dp$ are
  favored)~\cite{Taylor:2015rla}.

  The Auger Collaboration has found an indication of a possible
  correlation between UHECRs of $E > 10^{10.6}~{\rm GeV}$ and nearby
  starburst galaxies, with an {\it a posteriori} (post-trial) chance
  probability in an isotropic cosmic ray sky of $4.2 \times 10^{-5}$
  ($4\sigma$ significance)~\cite{Aab:2018chp}.  The energy threshold
  of largest statistical significance coincides with the observed
  suppression in the spectrum, implying that when we properly account
  for the barriers to UHECR propagation in the form of energy loss
  mechanisms~\cite{Greisen:1966jv,Zatsepin:1966jv} we obtain a self
  consistent picture for the observed UHECR horizon.  The TA
  Collaboration has reported that with their current
  statistics~\cite{Abbasi:2018tqo} they cannot make a statistically
  significant corroboration or refutation of the reported possible
  correlation between UHECRs and starburst galaxies. However, TA has
  recorded a statistically significant excess in cosmic rays, with
  energies above $10^{10.75}~{\rm GeV}$, above the isotropic
  background-only
  expectation~\cite{Abbasi:2014lda,Kawata:2015whq}. This is
  colloquially referred to as the ``TA hot-spot.'' The excess is
  centered at Galactic coordinates
  $(l,b) \simeq (177^\circ,50^\circ)$, spanning a region of the sky
  with $\sim 20^\circ$ radius. The chance probability of this hot spot
  in an isotropic cosmic ray sky was calculated to be
  $3.7 \times 10^{-4}$ (3.4$\sigma$ significance). The possible
  association of the TA hot-spot with the nearby ($3.4~{\rm Mpc}$ away)
  starburst galaxy M82 has not gone unnoticed~\cite{He:2014mqa,Pfeffer:2015idq}.
 
  We have seen that starburst galaxies can accommodate two of the main
  observables in UHECR physics: the nuclear composition and the
  distribution of arrival directions. We turn now to discuss the
  acceleration process in starburst superwinds, while
  exploring also whether this model can accommodate the shape of the source
  spectra. Before proceeding, we pause to note that very recently it
  was proposed that the UHECRs producing the TA hot-spot may be
  protons accelerated at sources in the Virgo Cluster (e.g. M87,
  $17~{\rm Mpc}$ away), which propagate towards the Earth along
  magnetic field filaments (of strength
  $\gtrsim 20~{\rm nG}$)~\cite{Kim:2019eib}. However, taking data at
  face value we can conclude that this proposal appears unlikely: {\it
    (i)}~as we have discussed above, a proton-dominated composition of
  UHECRs is disfavored by existing observations; {\it (ii)}~both the
  spectrum and the anisotropy observed by Auger and TA can {\it only} be
  accommodated if there is a steady source within $\sim 10~{\rm Mpc}$
  to account for the flux excess~\cite{Globus:2016gvy}. The latter
  reinforces the idea that the dominant source of the TA hot spot is
  the starburst galaxy M82. Another interpretation of the Auger
  anisotropy hints relies on UHECR acceleration in low-luminosity
  gamma-ray bursts~\cite{Zhang:2017moz}. However, if this
  were the case, the distribution of UHECRs would be isotropic in
  nature~\cite{Wang:2007ya}, or else correlate with the distribution of all nearby matter
  as opposed to a particular class of objects. It
  is noteworthy that when all sources beyond 1~Mpc (i.e. effectively
  taking out the Local Group) from the 2MRS catalog are included as
  part of the anisotropic signal in the analysis of~\cite{Aab:2018chp}
  the significance level reduces to $3\sigma$. Therefore, we can
  conclude that  the
  interpretation of the anisotropy signal in terms of low-luminosity
  gamma-ray bursts is disfavored by data.

  With the motivation loaded we can now look at the calculations.  The
  UHECR emission from starbursts is attributed to shock accelerated
  particles.  We describe the acceleration of these particles through
  the energy gain $g \equiv dE/dt$. We consider acceleration at
  superwind-embedded shocks in which the gain $g_{\rm SW}$ can be
  described by
\begin{equation}
g_{\rm SW}=\frac{\xi E}{T_{\rm cycle}},
\end{equation} 
where
\begin{equation}
T_{\rm cycle}=4\kappa\left(\frac{1}{u_1}+\frac{1}{u_2}\right)
\end{equation}
is the duration of each acceleration cycle,
\begin{equation}
\xi\sim\frac 4 3 (u_1-u_2)
\end{equation}
is the fractional energy gain per encounter,
\begin{equation}
\kappa=\frac{1}{3}R_{\rm L}\sim\frac{1}{3}\frac{E}{ZeB}
\end{equation}
is the diffusion coefficient, $R_{\rm L}$ is the Larmor radius, and
$u_1$ and $u_2$ are the upstream and downstream gas
velocities~\cite{Lagage:1983zz,Gaisser:1990vg}. For simplicity, we demand that any
two shocks do not propagate simultaneously. Studies of more general
set-ups, with shock correlation effects, are underway and will be
presented elsewhere~\cite{Soriano}. For typical superwind parameters
$u_2=u_1/4$ and
$u\equiv u_1\sim1.8\times10^3~\mathrm{km/s}$, the energy gain
\begin{equation}
  g_{\rm SW}^{(Z)}(B)=\frac{3}{20} Z e B u^2,
\label{Egain}
\end{equation}
produces a linear increase of energy as a function of time \begin{equation}
  \mathcal E(E_0,t_0,t)=E_0+g_{\rm SW}^{(Z)}(t-t_0),
 \end{equation}
for a fixed magnetic field.\footnote{The inferred value of $u$ from cold and
  warm molecular and atomic gas observations is  smaller
  than our fiducial value~\cite{Romero:2018mnb}. However, it is
  important to stress that the emission from the molecular and atomic
  gas most likely traces the interaction of the superwind with
  detached relatively denser ambient gas clouds~\cite{Heckman}, and as such it
  is not the best gauge to characterize the overall properties of the
  superwind plasma~\cite{Lacki:2013sda}.}  Thus, for an
accelerator of size $R_{\rm SW} \sim 8~{\rm kpc}$, the maximum energy is 
\begin{equation}
  E_{\rm max} \sim g_{\rm SW}^{(Z)} \, \Delta t \,,
 \label{pre-Hillas}
\end{equation}
where $\Delta t = t - t_0$. For a single shock, we
have $\Delta t = R_{\rm SW}/u$. Substituting this relation into
(\ref{pre-Hillas}) leads to the Hillas maximum
rigidity~\cite{Hillas:1985is}
\begin{equation}
  {\cal R}_{\rm H,max} \sim 10^9 \ (u/c) \left(\frac{B}{\mu{\rm G}} \right)\
    \left(\frac{R_{\rm SW}}{\rm kpc} \right) \, {\rm GV} \, .
\label{Hillas}
\end{equation}

To develop some sense of the orders of magnitude involved, we assume
that M82 and NGC 253 typify the nearby starburst population. The
magnetic field $B$ carries with it an energy density $B^2/(8\pi)$, and
the flow carries with it an energy flux $> uB^2/(8\pi)$. This sets a lower limit on
the rate at which the energy is carried by the out-flowing plasma,
\begin{equation}
  L_B \sim \frac{1}{8} \ u \ R_{\rm SW}^2 \ B^2 \,,
\label{eq:LIR}
\end{equation}  
and which must be provided by the source. The flux carried by the
outgoing plasma is a model dependent parameter, which can be characterized within an
order of magnitude. More concretely, $0.035 \lesssim L_B/L_{\rm IR} \lesssim
0.35$, where $L_{\rm IR} \sim 10^{43.9}~{\rm erg/s}$ is the infrared
luminosity~\cite{Heckman:1990fe}. The lower limit of $L_B$ corresponds to the estimate in~\cite{Heckman:1990fe} considering
a supernova rate of $0.07~{\rm yr}^{-1}$, whereas the upper limit  concur with 
the estimate in~\cite{Thompson:2006is}, and could be obtained
considering a supernova rate of $0.3~{\rm yr}^{-1}$~\cite{Bregman} while pushing other
model parameters to the most optimistic values. The relation
(\ref{eq:LIR}) yields a magnetic field strength in the range
\begin{equation}
15 \lesssim
B/\mu{\rm G} \lesssim 150 \, .
\label{B-range}
\end{equation}
Radio continuum and polarization observations of M82 provide
an estimate of the magnetic field strength in the core region of
$98~\mu{\rm G}$ and in the halo of $24~\mu{\rm G}$; see e.g. the
equipartition $B$ map in Fig.~16 of~\cite{Adebahr:2012ce}. Averaging
the magnetic field strength over the whole galaxy results in a mean
equipartition field strength of $35~\mu{\rm G}$. Independent magnetic
field estimates from polarized intensities and rotation measures yield
similar strengths~\cite{Adebahr:2017}. Comparable field strengths have
been estimated for NGC
253~\cite{Beck,Heesen:2008cs,Heesen:2009sg,Heesen:2011kj} and other
starbursts~\cite{Krause:2014iza}. Actually, the field strengths could
be higher if the cosmic rays are not in equipartition with the
magnetic field~\cite{Lacki:2013ry}. In particular, mG
magnetic field strengths have been predicted~\cite{Torres:2004ui} and
measured~\cite{McBride:2015} in the starburst core of Arp 220.  The
cosmic ray population in the starburst is dominated by the nearest
accelerators in time/space to the position of interest, thus breaking
a direct relation between average fields and mean cosmic ray
population~\cite{Torres:2012xk}.  Up to mG field strengths are consistent with the
gamma-ray and radio spectra in the gas-rich starburst cores of NGC 253
and M82~\cite{Paglione:2012ma}.  Besides, the field strength in the
halo of M82 and NGC 253 could be as high as
$300~\mu$G~\cite{DomingoSantamaria:2005qk,delPozo:2009mh,Lacki:2013nda}. In
our calculations we adopt the range given in (\ref{B-range}).

Substitution of (\ref{B-range}) into (\ref{Hillas}) leads to
$10^{8.9} \lesssim {\cal R}_{\rm H, max}/{\rm GV} \lesssim
10^{9.9}$. Taking this at face value, one would tend to interpret that
starburst superwinds {\it struggle} to accelerate light nuclei
($Z \lesssim 8$) up to the highest
observed energies~\cite{Matthews:2018laz}. Note, however, that in the case of multiple shocks the
time scale $\Delta t$ is not constrained by the ratio of the size of
the accelerator to the shock velocity, but rather by the lifetime of
the source $\Delta t \sim \tau$~\cite{Anchordoqui:1999cu,Anchordoqui:2018vji}. Then, for UHECRs experiencing the
effect of multiple shocks, the maximum rigidity is set by the Larmor
radius,
\begin{equation}
  {\cal R}_{\rm L,max} \sim 10^9 \ \left(\frac{B}{\mu{\rm G}} \right)\
    \left(\frac{R_{\rm SW}}{\rm kpc} \right) \, {\rm GV} \, ,
  \end{equation}
with an external constraint set by the energy loss. It is this that we now turn to suty. 

The final energy after the acceleration process for a fixed species $(A,Z)$ is given by the competition between the superwind acceleration and the possibility that a given nucleus suffers a photodisintegration and becomes a new species $(A',Z')$ after loosing one or several nucleons. A nucleus injected into the superwind at a time $t_0$ has probability $dP=f(t_0,t) dt$ to suffer a photodisintegration in the time in the interval $[t,t+dt]$, where
\begin{subequations}
\begin{equation}
f(t_0,t)=\frac{\mathcal F(t_0,t)}{\tau[\mathcal E(E_0,t_0,t)]} \,,
\label{eq:pdf-timea}
\end{equation}
\begin{equation}
\mathcal F(t_0,t)=\exp\left(-\int_{t_0}^t \frac{dt'}{\tau\bpar{\mathcal E(E_0,t_0,t')}}\right),
\end{equation}
\label{eq:pdf-time}
\end{subequations}
and $\tau(E)$ is the  mean free path for the nucleus at a given
energy. The accelerating nucleus will gain energy until it eventually
suffers a photodisintegration at a time $t$ distributed following (\ref{eq:pdf-timea}).

The photodisintegration rate depends on the energy density of the
ambient radiation field. This is governed by the spatial distribution
of photons, including both those from the cosmic microwave background
(CMB) and stellar radiation fields. For compact regions near the
galaxy core, starbursts exhibit an energy density in their stellar
radiation fields which may exceed (or be comparable to) that of the
CMB, but at the superwind scale $R_{\rm SW}$ starlight
is expected to have a negligible energy density compared to that of the
CMB~\cite{Anchordoqui:2007tn,Owen:2019msw}. For our
calculations, the contribution from the stellar radiation is
unimportant and therefore neglected.

\begin{figure}[tbp]
\begin{minipage}[t]{0.49\textwidth} \postscript{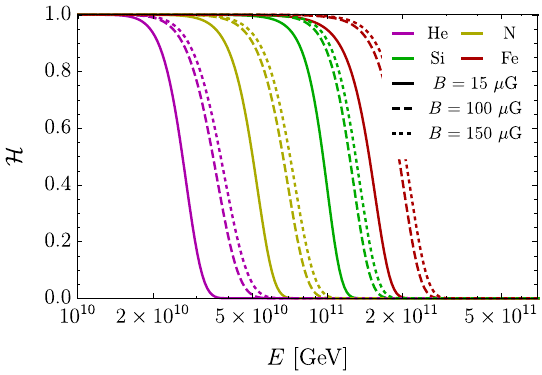}{0.99}
\end{minipage} \hfill \begin{minipage}[t]{0.49\textwidth}
\postscript{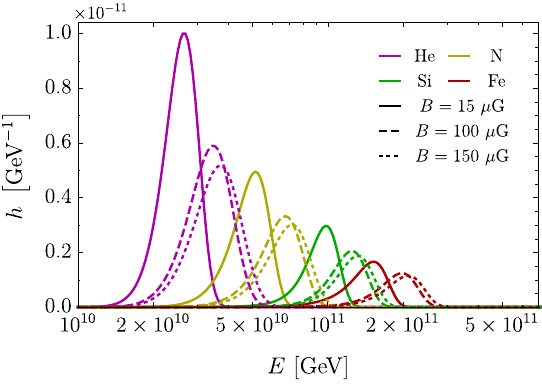}{0.99} \end{minipage}
\caption{Survival probability (left) and probability density for the energy at which the photodisintegration happens (right), for the four considered nuclei.}\label{fig:h}
\end{figure}

\begin{figure}[tbp]
\begin{minipage}[t]{0.49\textwidth} \postscript{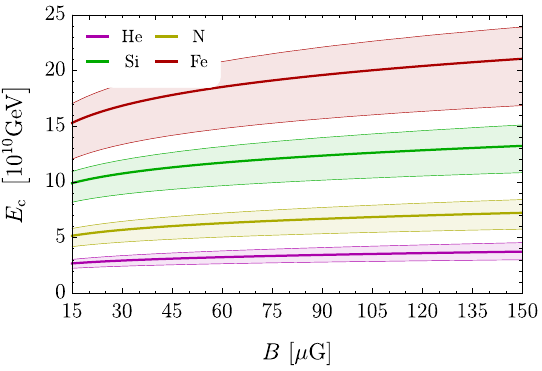}{0.99}
\end{minipage} \hfill \begin{minipage}[t]{0.49\textwidth}
\postscript{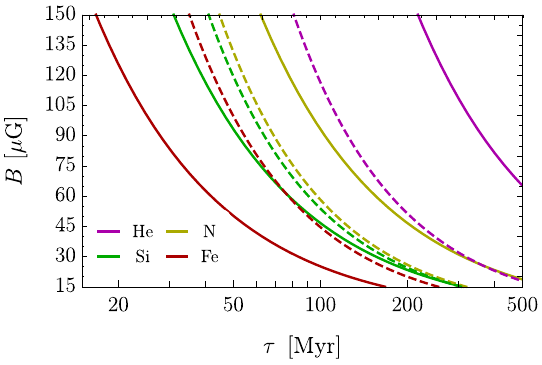}{0.99} \end{minipage}
\caption{Cutoff energy as a function of the magnetic field for the four considered nuclei (left), and parameter space for the source for a maximum energy of $10^{11}~{\rm GeV}$, and limits (dashed) imposed by CMB photodisintegration (right).}\label{fig:emax}
\end{figure}

In order to describe the energies that can be achieved through superwind acceleration, we consider the probability $dP=h(E_0,E)dE$ for the photodisintegration to happen at an energy in $[E,E+dE]$. Comparing with (\ref{eq:pdf-time}), the distribution for the final energy is
\begin{subequations}
\begin{equation}
h(E_0,E)=\frac{\mathcal H(E,E_0)}{g_{\rm SW}^{(Z)} \tau(E)},\label{eq:pdf-energya}
\end{equation}
\begin{equation}
\mathcal H (E_0,E)=\exp\left(-\int_{E_0}^E\frac{dE'}{g_{\rm sw}^{(Z)} \tau(E')}\right).
\end{equation}
\end{subequations}
The CMB mean free path is described as~\cite{Anchordoqui:2018qom}
\begin{equation}
\tau(E)=\left[\frac{c}{4\pi^2}\left(\frac{m}{\hbar c E}\right)^3\int_0^\infty\frac{J(\varepsilon)}{e^{\varepsilon /k T'(E)}-1}d\varepsilon\right]^{-1},
\end{equation}
where $T'(E)=2ET/ A m c^2$, $m$ is the proton mass, $T$ is the CMB temperature, 
\begin{equation}
J(\varepsilon)=\int_0^\varepsilon
\varepsilon'\sigma(\varepsilon')d\varepsilon' \,,
\end{equation}
and where $\sigma (\varepsilon')$ is the cross-section for
photo-disintegration by a photon of energy $\varepsilon'$ in the rest
frame of the nucleus.  The $\mathcal H$ functions have a decreasing sigmoid shape, as can be seen in Fig.~\ref{fig:h}. This allows to define a cutoff energy $E_c$ at the point of their largest decrease rate, which corresponds to the peak of the $h$ functions, shown in Fig.~\ref{fig:h} as well. This condition reads
\begin{equation}
E_c=\argmax_{E>E_0}\left(-\frac{d\mathcal H(E_0,E)}{dE}\right)=\argmax_{E>E_0}h(E_0,E).
\end{equation}
It can be rewritten in terms of the mean free path as 
\begin{equation}
1+\left.g_{\rm sw}^{(Z)} \frac{d\tau(E)}{dE}\right|_{E=E_c}=0,
\end{equation}
where the independence of $E_c$ on $E_0$ has been made explicit. The
values of the cutoff energy are shown in Fig.~\ref{fig:emax} for the
region of interesting  $B$-strengths.
The dispersion around the peak of $h$ suggests that particles of
energies above $E_c$ might be achieved at the source. In
Fig.~\ref{fig:emax} we show bands containing the $~68\%$ of the
probability (\ie such that $\mathcal H\sim0.16$ and
$\mathcal H\sim0.84$), which means that nuclei have around $16\%$
probability of reaching energies above (below) the top (bottom)
band. Further calculations show that nitrogen nuclei have a
probability of around $7\%$ of reaching energies above
$10^{10.95}~{\rm eV}$ and 1 \% above $10^{11}~{\rm GeV}$ for
$B \sim 150~{\rm \mu G}$. The sharp suppression of the probability function
with rising energy can accommodate a steeply falling spectrum if
sillicon-type nuclei are much less abundant than CNO-type nuclei. A
detailed study of this function with predictions on nuclear
composition to accommodate the observed spectrum on Earth will be presented
elsewhere~\cite{Soriano}.

The relevance of CMB photodisintegration at a certain source depends
on the interplay between the lifetime of the source and the mean free
path. For short living sources, the energies achieved will not be high
enough for the CMB to play a role, and the maximum energy for a
certain species will be determined solely due to the lifetime $\tau$
and the magnetic field $B$ as in (\ref{pre-Hillas}). This relation,
equivalently written as $B\propto E_{\rm max}/\tau$, allows to study
the parameter space $(\tau,B)$ that would allow to reach a certain
energy $E_{\rm max}$. Nevertheless, as the lifetime increases, the
available energies will be limited by photodisintegration processes
and, given a maximum energy, it will not always be possible to find a
pair $(\tau,B)$ able to provide such energy, since the CMB
interactions would produce a cutoff before that energy is reached. In
Fig.~\ref{fig:emax} we explore this parameter space for a maximum
energy of $10^{11}~{\rm GeV}$. The continuous lines follow
(\ref{pre-Hillas}), while the dashed lines define the regions (on the
right) that are not accessible due to the previous criterion. It can
be seen that on average no pair $(\tau,B)$ would be able to accelerate helium or nitrogen nuclei above that energy, while the opposite is true for silicon and iron nuclei.

In summary, we have re-examined the acceleration of UHECRs in starburst
superwinds endowed with multiple, non-simultaneous, propagating
shocks. Particles gain energy when they pass through the shock back
and forth after being scattered by the flowing plasma. To calculate the
maximum energy we must consider not only particles which are
accelerated by a single shock but also particles which undergo many
shock encounters, each of which further accelerates the
particles. There are two length scales which are important for
particle acceleration by multiple shocks: the mean free path for high
energy particles and the distance between shocks in the superwind. In
our approximation, only the first scale is relevant. We have shown
that the particle's maximum
energy is set by a balance equation driven by the source lifetime and UHECR interactions with the CMB. This gives specific
characteristics for the source emission spectra, providing a new
interpretation of the observed suppression in the UHECR spectrum.

Up until now, there were two competing classes of models to explain
the observed suppression in the energy spectrum.  The competing models
are: {\it (i)}~the GZK cutoff due to the UHECR interaction with the
CMB during propagation~\cite{Greisen:1966jv,Zatsepin:1966jv}, and {\it
  (ii)}~the disappointing model~\cite{Allard:2008gj,Aloisio:2009sj}
wherein it is postulated that the end-of-steam for cosmic accelerators
$\propto E_{\rm max}/Z$ is coincidentally near the putative GZK
cutoff. More concretely, conventional UHECR source models presuppose
that particle acceleration takes place at sites distributed similarly
to the matter distribution in the universe, with energy loss processes
during propagation leading to the observed flux suppression (GZK
cutoff). However, the most recent data seem to indicate that the
uppermost end of the cosmic ray energy spectrum is dominated by
nucleus-emitting- sources, possibly within the GZK horizon, for which
the upper limit of particle acceleration almost coincides with the
energy of the GZK suppression. In contrast to conventional
expectations, models in category {\it (ii)} suggest that the emission
of these sources would be characterized by a harder power-law spectrum
with the different mass components exhibiting a rigidity-dependent
maximum injection energy $E_{\rm max}/Z$ of a few EeV. Herein, we have
introduced an alternative possibility {\it (iii)} in which the maximun
energy is driven by GZK interactions, but as in {\it (ii)} the
observed suppression of the energy spectrum mainly stems from the
source characteristics rather than being the imprint of particle
propagation through the CMB.  Note that {\it (iii)} is markedly
different from {\it (ii)} because the maximum energy of acceleration
capability of the sources grows linearly in $Z$, while the energy loss
per distance traveled decreases with increasing $A$.

Class {\it (iii)} models have very particular predictions, which can
be easily distinguished from those in models of class {\it (ii)}. For
example, if the local distribution of sources dominates the spectrum
beyond the suppression, as suggested by anisotropy studies, our new
interpretation for the origin of the spectral cutoff explains {\it
  naturally} why the maximum energy observed on Earth coincides with
that expected from a uniform distribution of sources but with UHECR
nuclei propagating over cosmological distances. Moreover, the best fit
to the observed spectrum and nuclear composition yields a proton maximum
energy  $E_{\rm max}^p = E_{\rm max}/Z \sim 10^{9.5}~{\rm GeV}$~\cite{Aab:2016zth,Unger:2015laa}. This in turn gives a
maximum energy for CNO species of $E^{\rm CNO}_{\rm max} \sim
10^{10.5}~{\rm GeV}$, which is below the observed suppression in the
energy spectrum~\cite{Abraham:2008ru,Abraham:2010mj}, and therefore below the energy cutoff in the anisotropy analysis
of~\cite{Aab:2018chp}. Now, the typical values of the deflections of
UHECRs crossing the Galaxy are
\begin{equation}
  \theta \sim 10^\circ Z \left( \frac{E}{10^{10}~{\rm GeV}} \right)^{-1} \,,
\end{equation}
and therefore it is challenging to accommodate anisotropy patterns
with $Z \gtrsim 8$ nuclei~\cite{Anjos:2018mgr}. As we have shown, CNO species can be accelerated 
in starburst superwinds  to the maximum observed energies.

Altogether, this provides a compelling case demonstrating that there
  is strong evidence favoring UHECRs origin in starburst superwinds.

  \begin{center}
    {\bf Note Added}
    \end{center}
Additional support for the ideas discussed in this paper was presented
at the Conference. The Auger Collaboration reported the updated
    results of searches for anisotropies in the highest energy cosmic
    rays~\cite{Caccianiga}. With new data the significance of rejecting the isotropic hypothesis from
    a comparison with a starburst galaxies model has increased to
    reach $4.5\sigma$. The IceCube Collaboration reported the updated
    results of searches for neutrino emission from stacked catalogs of
    sources~\cite{Carver:2019jcd,Aartsen:2019kpk}. A catalog of 110 sources, comprising active galactic nuclei (including 
    blazars), starburst galaxies, and Galactic $\gamma$-ray sources was
    created using $\gamma$-ray data to select $\gamma$-bright sources
    that may produce neutrinos. The brightest neutrino source coincides with the brightest catalog
    source, the starburst galaxy NGC 1068. The significance of the source is 2.9$\sigma$
    after accounting for trials. In our model,  high-energy neutrino emission would be 
    expected from the starburst's core, where cosmic rays of energy
    $\lesssim E_0$ may experience an effective optical depth to
    hadronic interactions which is larger than unity. However, the
    neutrino emission would cutoff somewhat above $10^7~{\rm GeV}$, as entertained in~\cite{Loeb:2006tw}.

  \acknowledgments{We thank our colleagues of the Pierre Auger and POEMMA
    collaborations for some valuable discussions. This work has been
    supported by the by the U.S. National Science Foundation (NSF
    Grant PHY-1620661) and the National Aeronautics and Space
    Administration (NASA Grant 80NSSC18K0464).  Any opinions,
    findings, and conclusions or recommendations expressed in this
    material are those of the authors and do not necessarily reflect
    the views of the NSF or NASA.}

\section*{Appendix}

Consider a steady-state planar shock front endowed with a magnetic
field. Without loss of generality, the fluid velocity vector ${\bf u}$
and the magnetic field vector ${\bf B}$ can both be locally broken
down into components perpendicular to the shock front (designated by
the subscript $\perp$) and parallel to the shock front (designated by
the subscript $\parallel$). The Rankine-Hugoniot jump conditions for a
time-independent magnetohydrodynamic shock are found to be
\begin{equation}
  \llbracket  \rho u_{\perp } \rrbracket =  0 \,,
 \end{equation}
 \begin{equation}
\left \llbracket \rho u_{\perp}^2 + P + \frac{{B_{\parallel
      }}^2}{8\pi} \right \rrbracket = 0    \,,
\label{momentumcondition}
\end{equation}
  \begin{equation}
  \left \llbracket  \rho u_{\perp } u_{\parallel } - \frac{B_{\perp }
      B_{\parallel }}{4 \pi} \right \rrbracket = 0 \,,
  \end{equation}
\begin{equation}
  \left \llbracket \rho u_{\perp} \left(\frac{\gamma}{\gamma -1}
      \frac{P}{\rho} + \frac{u^2}{2} \right) -
    \frac{B_{\parallel }}{4 \pi} (B_{\perp} u_{\parallel } -
    B_{\parallel } u_{\perp } ) \right \rrbracket = 0 \,,
  \end{equation}
\begin{equation}
  \llbracket B_\perp u_\parallel - B_\parallel u_\perp \rrbracket =0 \,,
\end{equation}
\begin{equation}
 \llbracket B_\perp \rrbracket = 0 \, .
\end{equation}
where $\llbracket x \rrbracket \equiv x_1 - x_2$ expresses the jump
across the shock, $\rho$ is the mass density and $P$ the thermal
pressure, and where quantities measured just upstream of the shock are
designated by the subscript ``1'' and quantities measured just
downstream of the shock are designated by the subscript
``2''~\cite{deHoffmann}. Throughout we assume an ideal gas equation of
state and so the ratio of specific heats, $\gamma$, is considered a
constant parameter. It is straightforward to see that if the flow of
the gas is perfectly parallel to the field lines so that the shock
front is
oriented perpendicularly to them ({\it viz.} $u_\parallel =0$,
$u_\perp =u$, $B_\parallel = 0$,  and $B_\perp =
B$), then the momentum
jump condition (\ref{momentumcondition}) is the same as in the
hydrodynamical case. This makes sense on physical grounds, because
within this set up the ${\bf B}$ field exerts no net force on the gas,
so it is not surprising that the Rankine-Hugoniot jump conditions are
hydrodynamic in character~\cite{Anchordoqui:2018qom}. It is also straightforward to see
that when $B_\parallel \neq 0$ the jump condition for the momentum in
the perpendicular direction does depend on the magnetic field, which
provides an additional source of pressure. Moreover, the component of
the velocity parallel to the shock increases by a factor
\begin{equation}
  u_{\parallel 2} - u_{\parallel 1} = \frac{B_\perp}{4 \pi \rho
    u_\perp} (B_{\parallel 2} - B_{\parallel 1}) \, .
\end{equation}
For a perpendicular shock, in which $u_\parallel = 0$ and the magnetic
field is parallel to the shock front (i.e. $B_\perp = 0$), we can
solve for the density jump $\rho_2/\rho_1$ in terms of the ratio of the magnetic pressure to to the thermal
pressure
\begin{equation}
  \beta \equiv \frac{B_{\parallel 1}^2}{8 \pi P_1} \, ,
\end{equation}
and the Mach number
${\cal M}$ defined as the ratio of the unshocked gas speed to the upstream sound
speed. After discarding the trivial solution $\rho_2 = \rho_1$, the
Rankine-Hugoniot jump conditions  simplify
to the quadratic relation
\begin{equation}
  2 \ (2 -\gamma) \ \beta \ \left(\frac{\rho_2}{\rho_1} \right)^2 +
  \gamma \
  [(\gamma -1 ) \ {\cal M}^2 + 2 \ (\beta + 1)]
  \left(\frac{\rho_2}{\rho_1}\right) - \gamma \ (\gamma +1) \ {\cal M}^2 =
  0 \, .
  \end{equation}
Therefore, if the magnetic pressure is relatively insignificant (i.e. $\beta \ll
{\cal M}^2$) the change in density is approximately
\begin{equation}
 \frac{\rho_2}{\rho_1} = \zeta \left \{1 - \frac{4}{\gamma} \ \frac{\gamma
     + {\cal M}^2}{[2 + (\gamma -1 ) \ {\cal M}^2 ]^2} \ \beta \right\}
 \end{equation}
where $\zeta$ is the density ratio in the absence of a magnetic
field. The presence of a magnetic field thus tends to decrease the
density jump from what it would be in the absence of magnetism. As a
matter of fact, if we again examine carefully the quadratic equation for $\rho_2/\rho_1$,
we see that $\rho_2 > \rho_1$ only if
\begin{equation}
  {\cal M}^2 > 1 + \frac{2}{\gamma} \, \beta \, .
\end{equation}
This implies that the existence of a large magnetic field would allow supersonic
motions (${\cal M} > 1$) without the formation of shocks. In closing,
we stress once more that the model
under consideration herein is that of a parallel shock, for which  the
energy gain is given by (\ref{Egain})~\cite{Anchordoqui:2018qom,Protheroe:1998hp}.\footnote{Note
  that the
  conventions to identify parallel and perpendicular shocks followed
  throughout this paper are those in~\cite{Protheroe:1998hp} and reversed from those adopted
  in~\cite{Anchordoqui:2018qom}.}

\end{document}